# Spontaneous growth of diamond from MnNi solvent-catalyst using opposed anvil-type high-pressure apparatus

N. D. Zhigadlo*

*Laboratory for Solid State Physics, ETH Zurich, Otto-Stern-Weg 1, 8093 Zurich, Switzerland*

## Abstract

In this paper an overview of the application of opposed anvil-type high-pressure high-temperature apparatus for spontaneous growth of diamond crystals by solvent-catalyst technique is presented. The process makes use of a molten $Mn_{47}Ni_{53}$ catalyst to initiate the graphite-diamond transformation reaction. The pressure and temperature requirements to obtain reproducible spontaneous diamond crystallization were around 52.5-54 kbar and 1270-1320ºC. The crystals had well-shaped cubo-structured morphology with the {100} and {111} faces and grain sizes of 0.2-0.5 mm. This work can stimulate future experimental exploration of spontaneous diamond crystallization by using other solvent-catalyst metals.

* Corresponding author. Tel. +41 44 633 2249
E-mail address: zhigadlo@phys.ethz.ch

# 1. Introduction

Diamond is a superior material for many purposes due to its hardness, optical clarity, and resistance to chemicals, radiation, and electrical fields. However, most of natural diamonds are unsuitable for applications in electronics due to the presence of defects, impurities and poor structure. Nowadays, with the aid of high-pressure and high-temperature (HPHT) techniques for diamond formation [1,2] we are no longer dependent on natural diamond sources and can achieve properties well beyond those of natural diamond. Basically, the HPHT method uses equipment which tries to imitate the pressure and heat environment in which natural diamonds are found in the depths of the earth's crust. There are generally two methods for preparing synthetic diamonds: a direct and an indirect (or catalytic) one. The direct method involves only carbon and requires conditions in excess of ~130 kbar [3]. The indirect or catalytic conversion process can be carried out with pressure-temperature (P-T) conditions as low as 45 kbar and 1120°C, through the use of certain liquid metals or alloys. The ability of a transition metal to react with carbon increases with its number of electron vacancies in *d*-orbitals. Elements with no *d*-vacancies (e.g. Cu, Zn) are inert relative to carbon while those with many *d*-vacancies (e.g. Ti, V) are carbide formers. The most powerful elemental catalysts are transition metals with intermediate reactivities, such as Co, Fe, Mn, Ni and Cr [4]. These elements are major alloying components of catalysts used for the production of synthetic diamond under high pressure [5-11], while various nonmetallic solvents, such as the P-C system for example, are also promising candidates [12]. The mechanism of transformation of graphite materials into diamond in the presence of metal catalyst in the diamond-stable region of pressure and temperature is still not clear, although there have been many discussions [13]. The intent of this paper is to describe in detail the application of opposed anvil-type HPHT technique for spontaneous growth of diamond crystals. This method is less explored in the literature in comparison with the widely used cubic-anvil and belt-type techniques. By using an alternating sequence of MnNi solvent-catalyst and graphite layers sample assembly we determined the pressure and temperature regions suitable for growth of well-shaped cubo-structured diamond crystals with a typical size of 0.2-0.5 mm. These results reveal that the application of opposed anvil-type technique, in combination with solvent-catalyst metals can play an important role in further exploring the possibilities to obtain crystalline diamond.

## 2. Experimental details

The HPHT experiments were performed in the opposed anvil-type high-pressure apparatus shown in Fig. 1. The pressure generation mechanism is based on the compression and confinement of a solid medium (lithographic stone) between special working surfaces of a pair of opposed anvil-type dies (Fig. 1, left panel). The assembled high-pressure module presented on the right panel of Fig. 1, takes the simple forward motion of a hydraulic jack and converts it into the inward motion of the anvils. During pressurization, the lithographic stone squeezes out into the spaces between the anvils until the friction between the pressure medium and the anvils balances the pressure generated inside the sample assembly. The pressure applied to the sample was estimated from the load-pressure relation obtained from separate experiments, employing Bi and PbSe as calibrants. The temperature of the sample was determined by the pre-calibrated relation between the applied electrical power and the measured temperature in the cell.

The upper and the lower panels of Fig. 2 show a schematic illustration of the cross-sectional view of the high-pressure cell and sample arrangement. The furnace consists of a graphite tube that contacts two of the anvils through Mo, Ta, Fe and Cu conductive metals. The heater is isolated from the lithographic stone by disks made of a 50%NaCl/50%$ZrO_2$ powder mixture. The boron nitride (BN) crucible contains the sample and prevents it from reacting with other parts of the sample assembly. Typical growth procedures last less than one hour. All the retrieved HPHT products were treated in a hot mixture of $H_2SO_4$ and $HNO_3$ acids to remove residual graphite and solvent catalyst metals. The general morphology and dimensionality of the grown crystals was observed by an optical microscope (Leica M 205C).

## 3. Results and discussion

The diamond crystal growth at HPHT conditions is a process of dissolution of graphite in the molten catalyst to achieve the transition of graphite into diamond under the influence of a metal catalyst [14]. For the efficient spontaneous diamond crystallization from solvent-catalysts, the detailed knowledge of the P-T parameters is required. In this work the melting temperature line for the eutectic composition of $Mn_{47}Ni_{53}$ solvent-catalyst (dashed red line 2 in Fig. 3) was estimated based on the liquid-solid Mn-Ni equilibrium diagrams at hydrostatic pressures up to 100 kbar, as first obtained by Ershova *et al* [15] and later on summarized by Gokcen [16]. A notable aspect of these diagrams is the faster increase of melting point of pure Mn with increasing pressure, with respect to pure Ni. The optimal region of diamond formation in the Mn-Ni-C system is bounded by the eutectic melting temperature line and the graphite-diamond equilibrium line. With this we can estimate the minimum P-T conditions required for the diamond crystallization, which seem to be very close to the intersection between the two lines (i.e., lines 1 and 2 in Fig. 3). Following these guidelines several HPHT synthesis experiments were carried out at a pressure of 52.5-54 kbar and at a temperature range of 1270-1320°C. A reactive mixture was composed by alternating layers of graphite disks and $Mn_{47}Ni_{53}$ catalytic powder (Fig. 2, upper panel). Disks with a cavity made of commercial isotropic ultra-fine grain polycrystalline graphite rod (Poco EDM-3) were filled with fine powdered MnNi solvent-catalyst and stacked in the BN crucible. Various thicknesses of stacked alternating layers were tested and the best results were obtained for a 1-mm MnNi and 1.5-mm graphite sequence of layers. In a typical growth run, a pressure of 52.5-54 kbar is applied at room temperature. While keeping the pressure constant, the temperature is ramped up within 10 minutes to the maximum value of 1270-1320°C, is kept constant for 2.5-5 mins and then the cell is cooled to room temperature in 5-7 mins. The high pressure was maintained constant throughout the synthesis and was removed only after the cell was cooled to room temperature. We note that one of the important factors influencing the temperature characteristics of the reactive cell and thus the overall growth process is the design of the heater and of the adjacent parts of the power supply. In practice, usually in cubic-anvil type and belt-type apparatus, the graphite heater in the form of a cylindrical tube with direct contact to the anvils is used [17]. In such design, the heat transfer at high temperature leads to a situation where a small increase in temperature demands a large increment in heater power. In the present work the contact of graphite heater with the opposed anvils was carried out through metal pieces, i.e. two steel cylinders with a diameter of 4 mm and Cu, Ta and Mo plates. This construction leads to a considerable reduction in the demanded heating power and also allows for lowering the relaxation time down to 10 seconds, where, the relaxation time defined as the time needed by the reaction cell to respond to a power change and achieve the steady state. Altogether, this design permits more controllable growth process.

Dozens of diamond crystals with sizes of 0.2-0.5 mm were obtained from each experimental run. Most of the crystals were yellow in color, but occasionally some were still covered with black residuals of graphite and/or MnNi solvent. Typical well-shaped cubo-structured morphologies with the {100} and {111} faces and without any melting holes can be clearly seen in Fig. 4. A few strip-shaped diamond crystals were also formed in our experimental conditions. The appearance of such morphology has been reported in a nitrogen-rich system [18, 19] and usually most of the obtained diamond stripes elongate or extend only along the {111} direction, with no obvious {100} faces. Probably, in our case the graphite-diamond conversion is also partially affected by nitrogen incorporation, although we have not yet clear experimental proofs. These results demonstrate that an opposed anvil-type device in combination with the metal-solvent concept can be successfully used to further explore spontaneous diamond crystallization.

Not going into details of the possible mechanism of the diamond growth, since it requires more systematic investigations, we note here the importance of the starting raw carbon for the diamond synthesis. It is well known that the synthesis of diamond under HPHT occurs through a dissolution-precipitation process, i.e. dissolution of raw carbon into a metallic catalyst and its precipitation as diamond [14]. The analysis of HPHT experiments where the different types of graphite were used reveals that the degree of graphitization, which corresponds to the ability of a carbonaceous material to rearrange its atomic structure towards the hexagonal graphite structure influences significantly the crystal growth process and the morphological characteristics of diamond formed [20-22]. For example, Skury *et al* [21,22] investigated the influence of the degree of graphitization on the diamond growth through HPHT treatment of a reactive mixture composed, as in our case, by alternating layers of graphite powder and $Mn_{58}Ni_{42}$ alloy catalyst and they too found that the growth is significantly enhanced by using graphite with high level of graphitization. With regard to the crystalline structure, the Poco EDM-3 synthetic graphite used in this study has a typical hexagonal structure consisting of a succession of layers parallel to the basal plane of hexagonally linked carbon atoms with high level of crystallinity. Thus, it is reasonable to assume that along with the processing parameters the quality of the used graphite accounts for the growth of well-shaped diamond crystals.

## 4. Conclusions

This paper describes the application of an opposed anvil-type high-pressure high-temperature technique for the spontaneous growth of diamond crystals by using a MnNi-based solvent-catalyst system. We could identify the lower pressure and temperature region of the diamond formation at 52.5 kbar and 1270°C. The used alternating layers sample assembly yields of 0.2-0.5 mm sizes crystals with well-shaped cubo-structured morphologies. The results presented in this work have the potential to be extended to the growth of diamond crystals by using other metal solvents.

## Acknowledgment

The author is indebted to N. Nikolaev for his expert assistance in setting up the high-pressure apparatus and for conducting the high-pressure experiments.

**Figures and captions**

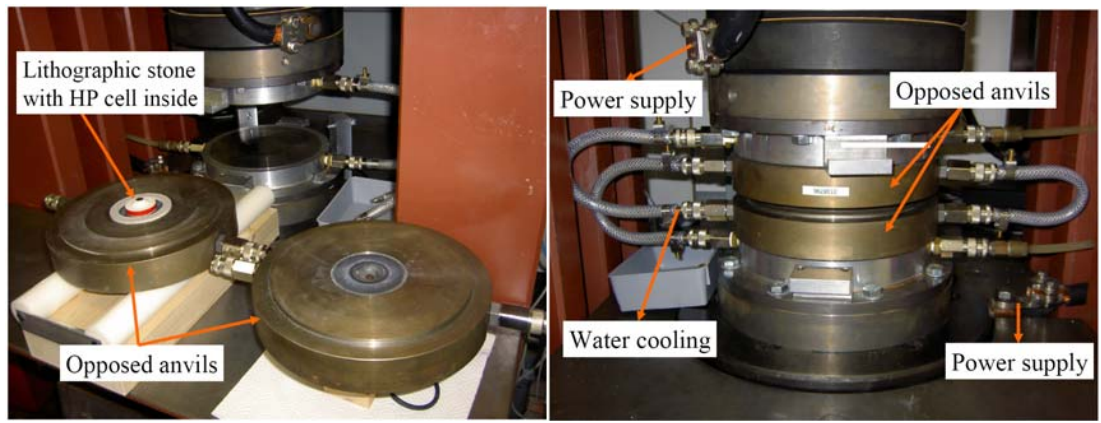

**Fig. 1.** Left panel: Opposed anvil-type, high-pressure device, with lithographic stone as a pressure medium. Right panel: Overview of assembled high-pressure module of opposed anvil-type high-pressure device. The pressure generation mechanism is based on the compression and confinement of a solid medium between special working surfaces of a pair of opposed-anvil type dies. The module takes the simple forward motion of the hydraulic jack and converts it into the inward motion of the anvils. During pressurization, the pressure medium squeezes out into the spaces between the anvils until the friction between the pressure medium and the anvils balances the pressure generated inside the sample assembly.

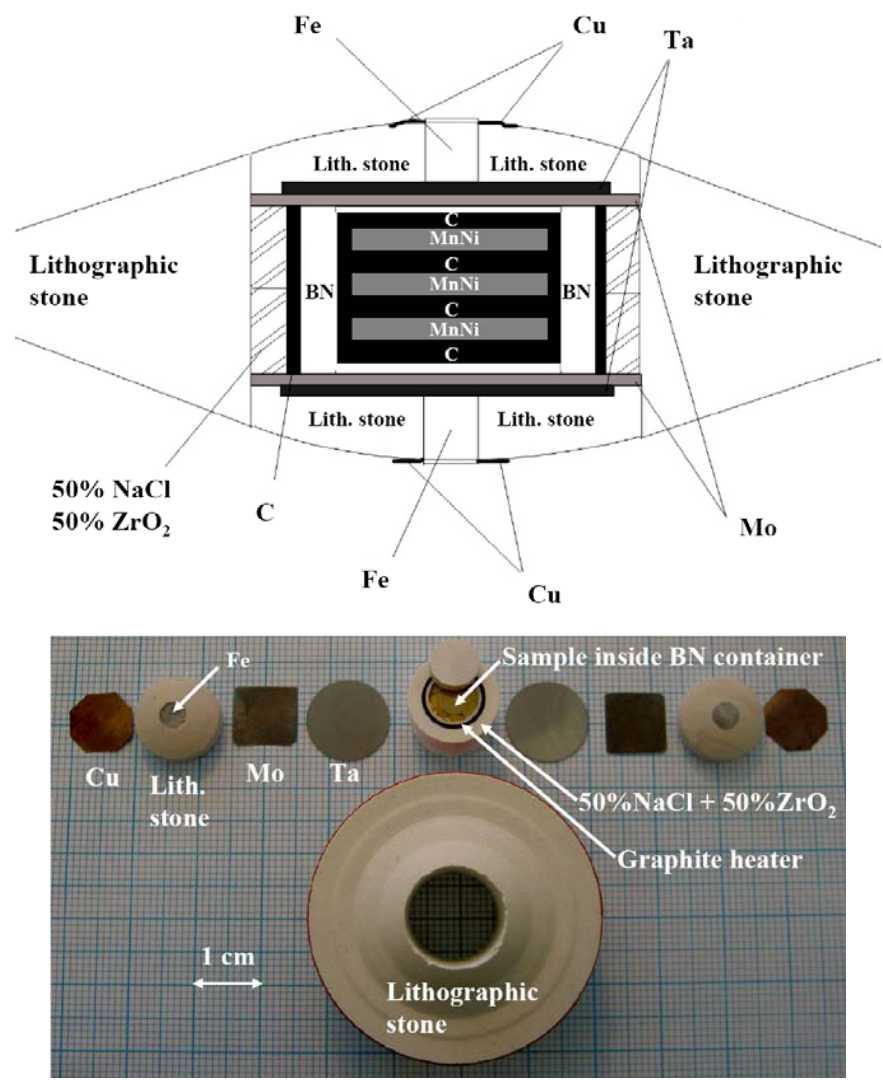

**Fig. 2.** Schematic illustration of the cross-sectional view of the high-pressure cell and sample arrangement. The furnace is a graphite tube that makes contact with two anvils through a Mo, Ta, Fe and Cu conductive material. The BN crucible contains the sample and prevents it from reacting with other parts of the sample assembly. Graphite and MnNi solvent-catalyst were stacked in the reaction cell as alternate layers.

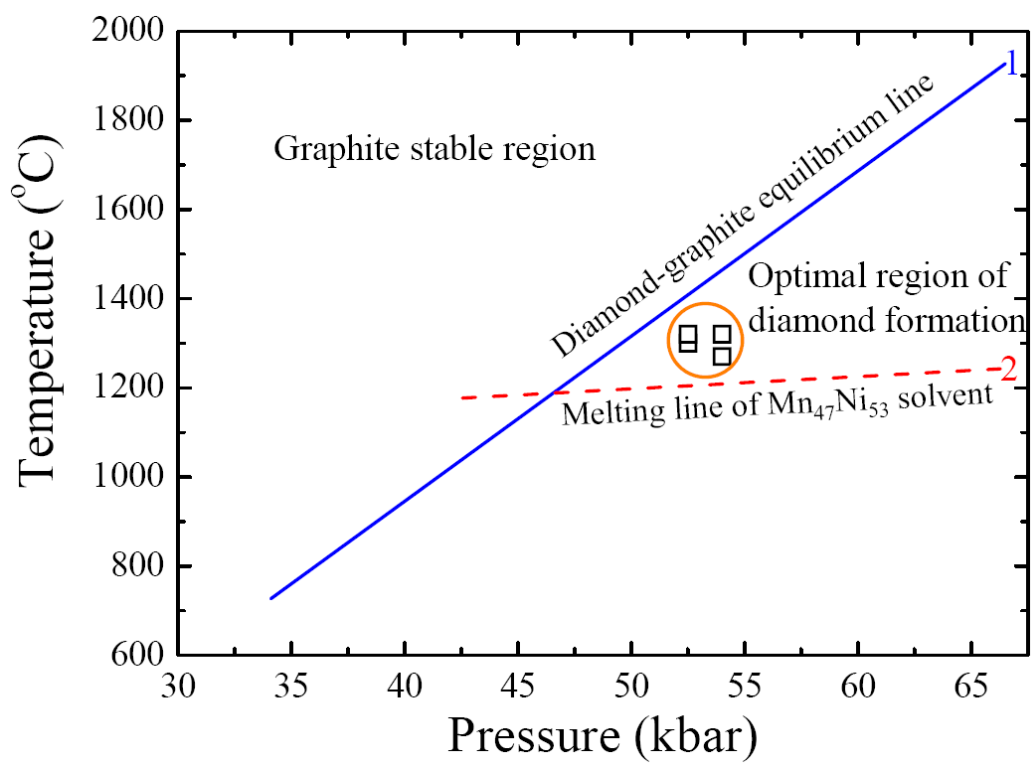

**Fig. 3.** Schematic pressure-temperature phase diagram for the diamond synthesis process. The orange circle shows the optimal region for the diamond crystallization in the Mn-Ni-C system. Square symbols are the experimental points. The graphite-diamond equilibrium line (1) was plotted according to Bundy *et al* [14]. The melting curve for the $Mn_{47}Ni_{53}$ solvent-catalyst (2) was estimated from to the data presented in ref. [15, 16].

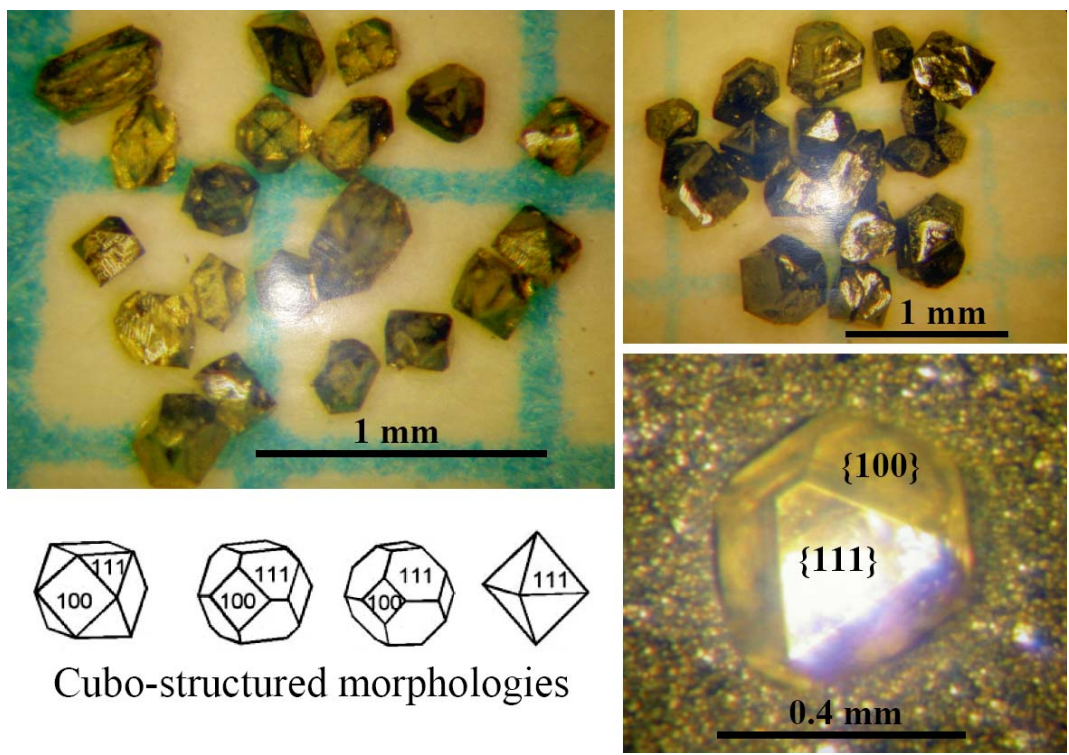

**Fig. 4.** Optical images of diamond crystals obtained from $Mn_{47}Ni_{53}$ solvent catalyst at a pressure of 52.5-54 kbar and a reaction temperature of 1270-1320°C. The crystals had well-shaped cubo-structural morphology with {100} and {111} faces. A few strip-shaped crystals are also formed in our experimental conditions.